\documentclass[fleqn,twocolumn]{wlscirep}

\usepackage[utf8]{inputenc}
\usepackage[T1]{fontenc}

\usepackage{amssymb} \usepackage{color,graphicx} \usepackage{amsmath}
\usepackage{amsbsy} \usepackage{amsthm} 
\usepackage{bm} \usepackage{float} 
\usepackage{placeins,cals}
\usepackage{setspace,ulem}

\usepackage[font=small,labelfont=bf,
   justification=justified,
   format=plain]{caption}

\newcommand{\bq}{\begin{equation}} \newcommand{\eq}{\end{equation}}
\newcommand{\bqali}{\begin{equation}\begin{aligned}} \newcommand{\eqali}{\end{aligned}\end{equation}}
\newcommand{\D}{\operatorname{d}}

\usepackage{braket}

\newcommand\rC{r_\text{\tiny C}}
\newcommand{\kB}{k_\text{\tiny B}}

\title{{Present status and future challenges of non-interferometric tests of collapse models}}

\author[1]{Matteo Carlesso}
\author[2]{Sandro Donadi}
\author[2,3]{Luca Ferialdi}
\author[1]{Mauro Paternostro}
\author[4]{Hendrik Ulbricht}
\author[2,3,*]{Angelo Bassi}
\affil[1]{Centre for Theoretical Atomic, Molecular, and Optical Physics,
School of Mathematics and Physics, Queens University, Belfast BT7 1NN, United Kingdom}
\affil[2]{Istituto Nazionale di Fisica Nucleare, Trieste Section, Via Valerio 2, 34127 Trieste, Italy}
\affil[3]{Department of Physics, University of Trieste, Strada Costiera 11, 34151 Trieste, Italy}
\affil[4]{School of Physics and Astronomy, University of Southampton, SO17 1BJ Southampton, United Kingdom}
\affil[*]{abassi@units.it}

\begin{document}

\thispagestyle{empty}

\begin{abstract}
The superposition principle is the cornerstone of quantum mechanics, leading to a  variety of genuinely quantum effects.
Whether the principle applies also to macroscopic systems or, instead, there is a progressive breakdown when moving to larger scales, is a fundamental and still open question. Spontaneous wavefunction collapse models predict the latter option, thus questioning the universality of quantum mechanics. 
Technological advances allow to challenge collapse models and the quantum superposition principle more and more with a variety of different experiments. Among them, non-interferometric experiments proved to be the most effective in testing these models. We provide an overview of such experiments, including cold atoms, optomechanical systems,  {X}-rays detection, bulk heating as well as comparisons with cosmological observations. We also discuss  avenues for future dedicated experiments{,} which aim at further testing collapse models and the validity of {quantum mechanics}. \end{abstract}

\flushbottom
\maketitle

Quantum mechanics radically changed our understanding of Nature. 
The superposition principle allows for the preparation of quantum systems in coherent superpositions of distinguishable physical configurations. This challenges our classical intuition according to which objects can only be in a one definite physical state at a time.
After {almost} one hundred years of experimental endeavours, the validity of the superposition principle at the microscopic scale is beyond questioning. It has led to an unprecedented understanding of the behaviour of matter and light and to the development of several quantum technologies, such as the laser and the transistor, which are now part of our everyday life. 

Despite such success, we face a puzzling situation at the macroscopic scale: we do not experience quantum superpositions, although quantum mechanics does not set any explicit upper bound to the size that such {superpositions} can have. One {possible explanation for the lack of observation of macroscopic quantum superpositions} is that the superposition principle progressively breaks down when moving from the microscopic to the macroscopic world~\cite{penrose1996gravity,Adler:2004wc,Leggett871,weinberg2017trouble}.

 In this regard, spontaneous wavefunction collapse models --- or simply collapse models --- provide a consistent  phenomenological framework for the breakdown of quantum superpositions. The collapse mechanism becomes stronger with the size and complexity of a given system, so that, while the microscopic world is  quantum mechanical, the macroscopic world is  classical. The collapse dynamics, which is controlled by few parameters,  differs from the standard quantum dynamics{. The} differences can be verified experimentally, and we have recently witnessed an increasing effort in 
 placing strong experimental bounds on the value of their parameters.

There are essentially two methods to test collapse models. The most direct approach is to perform interferometric experiments, aiming at detecting quantum superpositions (or the lack thereof) with larger and larger objects~\cite{arndt2014testing}. The alternative approach is to  {conduct} non-interferometric experiments, where the possible violation of the  superposition principle is tested indirectly through various side-effects of the collapse dynamics. 

Despite their immediacy, interferometric experiments 
become significantly harder to perform when the size of the  system to test grows. 
Non-interferometric experiments 
are relatively easier as they do not require  {one to} prepare  the system in a quantum superposition.  {Instead, they require} the precise monitoring of quantities such as the position or the energy. 

This review addresses non-interferometric experiments and their ability to provide bounds in the parameter space of  
two of the most important collapse models, the Continuous Spontaneous Localization model~\cite{pearle1989combining,ghirardi1990markov} (CSL) and the Di\'osi-Penrose (DP) one~\cite{diosi1987universal,penrose1996gravity}. 

\begin{figure*}[t]
    \centering
    \includegraphics[width=0.9\linewidth]{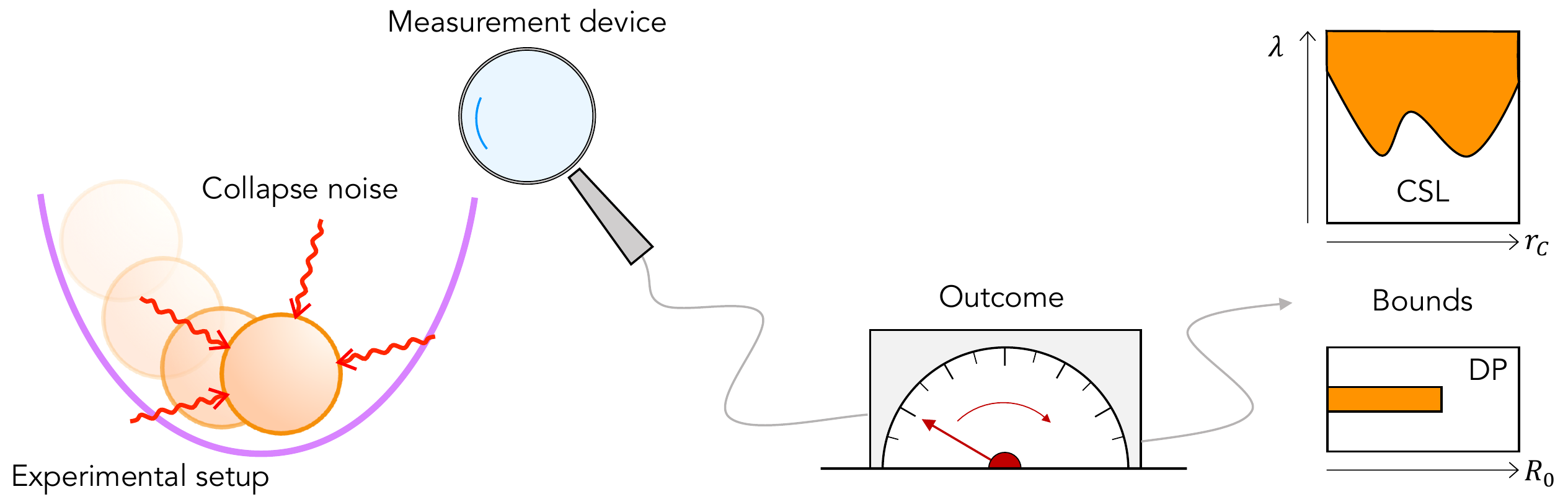}
    \caption{{Testing collapse effects using a typical non-interferometric setup}. The system (shown as an orange sphere) evolves as described by its quantum mechanical dynamics (for example, there can be a potential, represented by the purple line). The collapse noise (here identified by the red arrows) will modify such dynamics, thus providing predictions which are different from those of quantum mechanics. A suitable measurement device (sketched as a magnifying glass) aims at detecting such a difference. The measurement outcomes are then used to draw the experimental upper bounds on the CSL model and lower bounds on the DP model, which are shown in Fig.~\ref{fig:csl} and Fig.~\ref{fig:dp} respectively. 
        }
    \label{fig:measurement}
\end{figure*}

\begin{figure}[ht!]
    \centering
    \includegraphics[width=\linewidth]{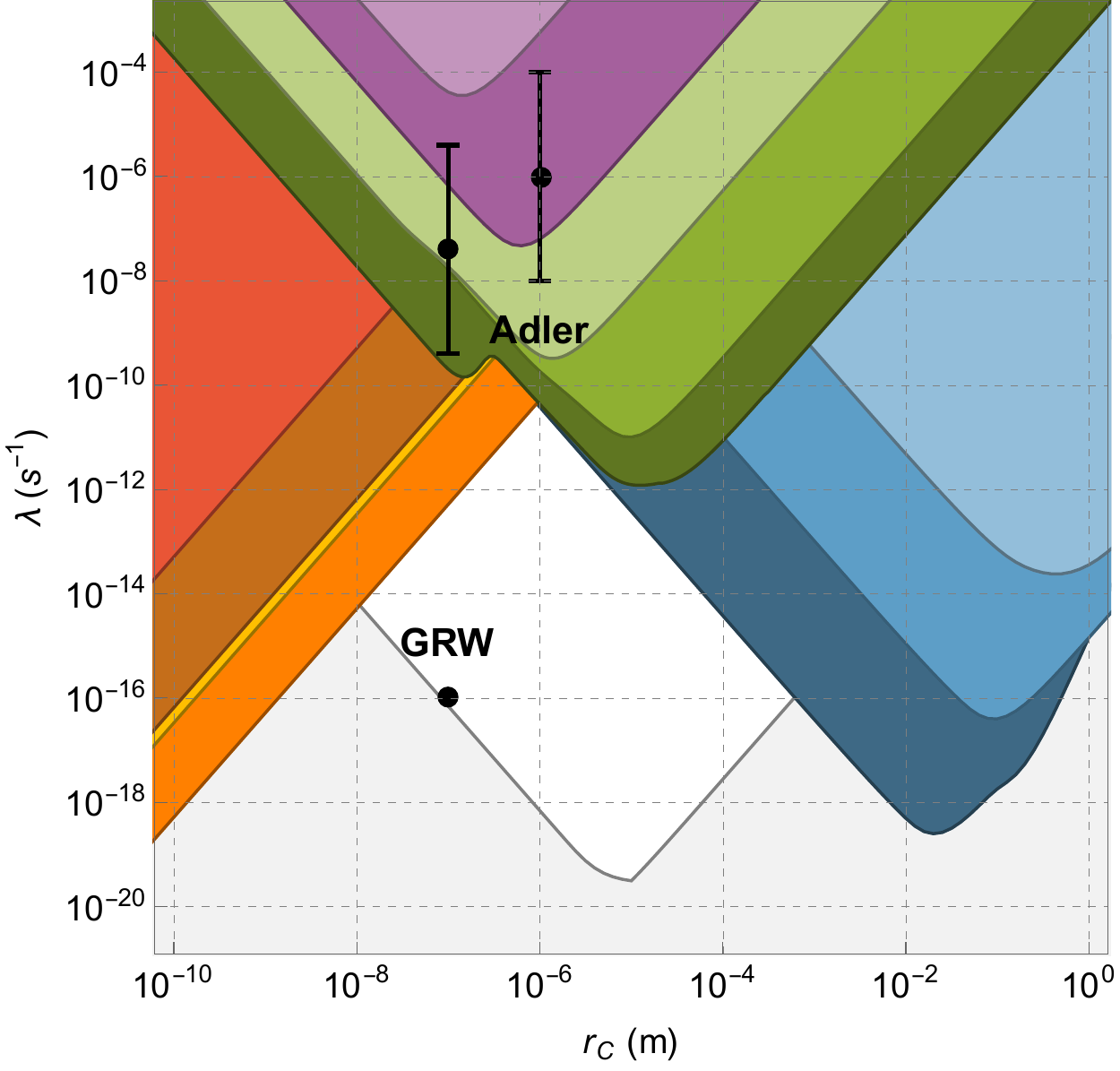}
    \caption{Exclusion plot for the CSL parameters $\lambda$ and $\rC$ from non-interferometric tests. The colored areas correspond to experimentally excluded regions. The green-colored regions are from cantilever-based experiments with masses of $\sim10$ ng~\cite{vinante2016upper} (light green), $\sim100$ ng~\cite{vinante2017improved} (green), and multilayer structures~\cite{vinante2020narrowing} (dark green). Blue areas are obtained from gravitational wave detectors~\cite{carlesso2016experimental,helou2017lisa,carlesso2018non}: AURIGA (light blue), LIGO (blue) and LISA Pathfinder (dark blue). Purple areas are from optomechanical systems levitating in  a linear Paul trap~\cite{pontin2020ultranarrow} and a magnetic trap~\cite{zheng2020room}. The orange area is from spontaneous X-ray emission tests~\cite{Donadi:2021tq}. The yellow area is from phonon excitation in the CUORE experiment~\cite{pobell2007matter,adler2018bulk}. The brown area is from the heating rate of Neptune~\cite{adler2019testing}. The red area is drawn from cold-atom experiments~\cite{bilardello2016bounds}. The theoretical values proposed by GRW~\cite{ghirardi1986unified} and {the ranges proposed by} Adler~\cite{adler2007lower} are shown respectively as a black dot and black dots with bars that indicate the estimated range.  {Finally, the light grey area is excluded not from experiments but from the requirement that macroscopic superpositions do not persist in time, which is the main motivation behind collapse models. Specifically, the (relatively arbitrary but reasonable) requirement adopted here is that a graphene disk of radius 10\,$\mu$m, approximately the smallest visible size for a human eye, collapses in 0.01\,s, which is about the  time resolution of the human eye~\cite{torovs2017colored}.} The white area is yet to be explored. 
    }
    \label{fig:csl}
\end{figure}

\begin{figure}[t!]
    \centering
    \includegraphics[width=\linewidth]{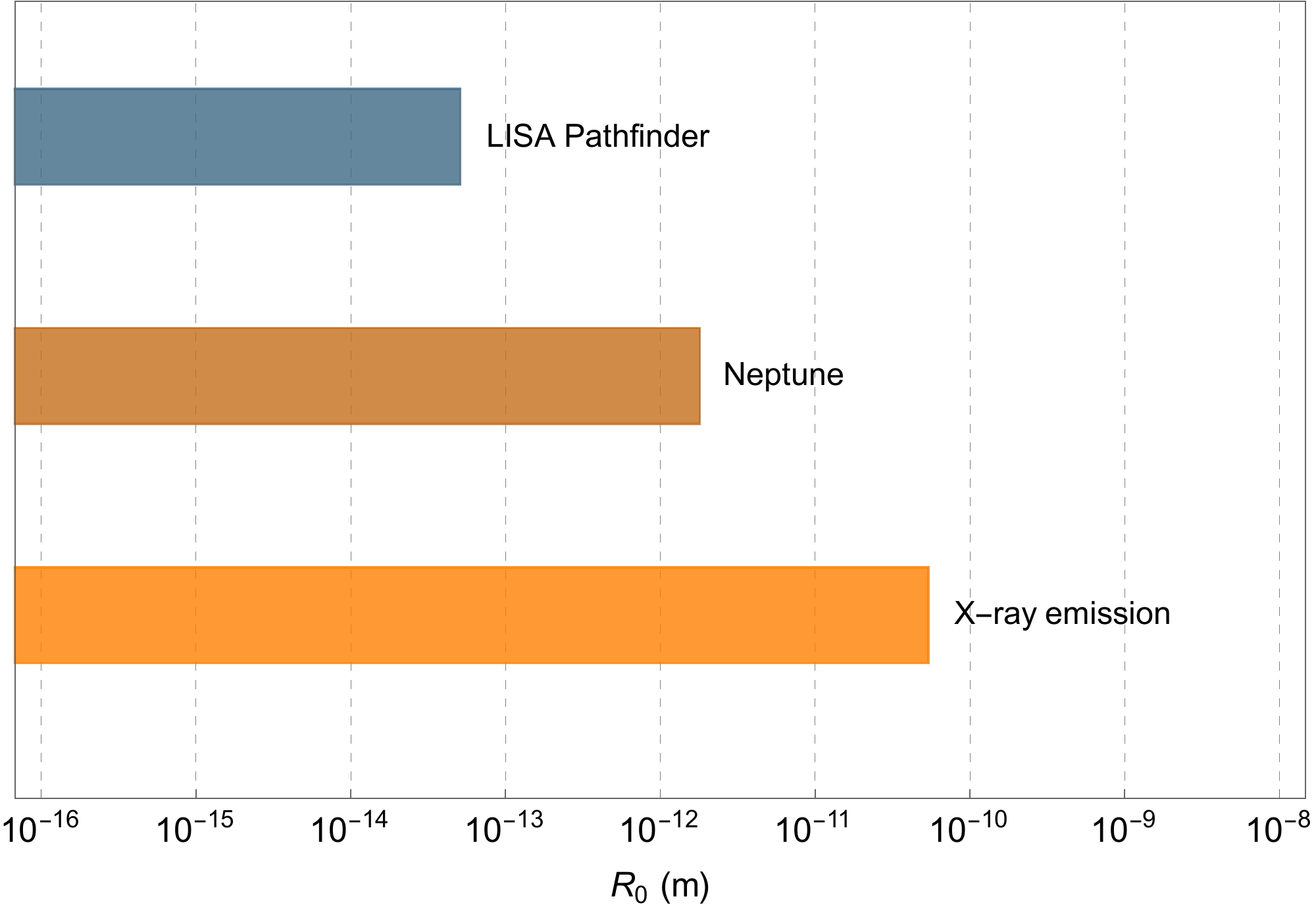}
    \caption{Exclusion plot for the DP parameter $R_0$ from non-interferometric tests. The colored areas correspond to experimentally excluded values of $R_0$. The blue bound is from LISA Pathfinder~\cite{helou2017lisa}, the brown area is from the heating rate of Neptune~\cite{adler2019testing}, the orange one is from X-ray emission tests~\cite{donadi2020underground}. 
    }
    \label{fig:dp}
\end{figure}

\section{Theoretical framework {of collapse models}}

Collapse models provide a mathematically and physically consistent dynamical framework, where quantum superpositions and  wavepacket reduction are combined.
This is achieved by embedding in the Schr\"odinger equation 
the mechanism responsible for wavepacket reduction --- the not-better-specified collapse of the wavefunction to a definite state upon a measurement according to the standard formulation of quantum theory.
{Such a mechanism has two features, the first} is nonlinearity, which is needed to break the superposition principle. The second one is stochasticity, which allows to recover quantum indeterminacy. 

{In order to avoid {superluminal} signaling, non-linear and stochastic terms must be blended carefully~\cite{Adler:2004wc,gisin1989stoch}.} This  {yields} a well specific structure of the dynamical equation. For the models considered in this review, and using the It\^o formalism for stochastic differential equations~\cite{arnold1971stoch}, such a {dynamical} equation reads~\cite{ghirardi1990markov,PhysRevLett.73.1,bassi2013models}
\bqali
\label{csl}
&\D \ket{\psi_t}= \left[- \frac{i}{\hslash} \hat{H} \D t+ \int \D^3 {\bf x}\,\left(\hat{M}({\bf x})-\braket{ \hat{M}({\bf x}) }_t \right)\, 
\D W_t({\bf x})\right.\\
&\left.
-  \frac{1}{2}\int \D^3{\bf x}\D^3{\bf y}\,{\mathcal D}({\bf x}-{\bf y})\prod_{{\bf q}={\bf x},{\bf y}}
\left(\hat{M}({\bf q})-\braket{ \hat{M}({\bf q}) }_t \right)
\,\D t  \right] \ket{\psi_t}{,}
\eqali
{where $\hslash$ is the reduced Planck constant.} 
{The first term on the right-hand side is the standard quantum contribution as encoded by the system Hamiltonian $\hat H$. The second and third terms describe the stochastic non-linear collapse process weighted by the}  mass density operator $\hat{M}({\bf x})$, which {ensures} that the wavefunction is progressively localized in space.  {The collapse process is} driven by the Brownian noise $W_t({\bf x})$ with spatial correlation equal to $\cal{D}({\bf x}-{\bf y})$, and by the non-linear contribution to the dynamics $\braket{\hat M({\bf q})}=\braket{\psi_t|\hat M({\bf q})|\psi_t}$. 

{{It is worth stressing that Eq.~\eqref{csl} is built in a way that the statistical operator $\hat\rho_t=\mathbb{E}[\ket{\psi_t}\bra{\psi_t}]$ (where $\mathbb E$ is the stochastic average with respect to the noise) 
obeys the Lindblad equation
\begin{equation} \label{CSLme}
\frac{\D}{\D t}\hat\rho_t  =  - \frac{i}{\hslash} \left[ \hat{H},
\hat \rho_t \right]  +   \int \D^3{\bf x}\D^3{\bf y}\,{\cal D}({\bf x}-{\bf y})  [\hat{M}({\bf x}), [\hat{M}({\bf y}) , \hat \rho_t]].
\end{equation} 
In contrast to the collapse-modified Schr\"odinger equation, the dynamics in Eq.~\eqref{CSLme}  {are} linear. {{This} forbids the possibility of {superluminal} signaling, in spite of the fact that collapse is a non-local process~\cite{gisin1989stoch}. 
}
Although the collapse of the wavefunction is now hidden, Eq.~\eqref{CSLme} is easier to solve when computing the evolution of expectation values of operators. }

Notice that the  dynamics resulting from Eq.~\eqref{csl}, although not unitary,  {are} norm-preserving {and} also embed an amplification mechanism: the collapse rate of an object scales roughly with its size.
{Consequently, one} can  set extremely small values for the collapse rate for microscopic systems, thus effectively recovering the standard unitary quantum evolution. In turn,  the amplification mechanism implies a large collapse rate for macroscopic systems, which remain well localized in space, thus retrieving classical mechanics. In particular, when a microscopic system interacts with a macroscopic measuring device, the collapse dynamics makes sure that  {the} outcomes at the end of the measurement {are definite}, which are distributed according to the Born rule. In this framework, the Born rule is not assumed but derived~\cite{bassi2013models}.

The two most studied collapse models are the CSL and the DP model, which are both described by Eq.~\eqref{csl} with different  choices of the correlator $\mathcal{D}({\bf x}-{\bf y})$. The CSL model assumes a Gaussian correlator $\mathcal{D}_\text{\tiny CSL}({\bf x}-{\bf y}) = \frac{\lambda}{m_0^2}\exp\left(-{|{\bf x}-{\bf y}|^2}/{4 \rC^2}\right)$ ($m_0$ is the mass of a nucleon), characterized by  two phenomenological parameters: the collapse rate $\lambda$, which sets the strength of the collapse for a single nucleon, and the  length $\rC$ beyond which spatial superpositions are suppressed. 
The value proposed by Ghirardi, Rimini, and Weber~\cite{ghirardi1986unified} (GRW) for the collapse rate is $\lambda=10^{-16}$\,s$^{-1}${, which guarantees an effective collapse only for macroscopic systems}, {whereas} Adler~\cite{adler2007lower} proposed the larger values   {$\lambda=4\times 10^{-8\pm2}$\,s$^{-1}$ at $\rC=10^{-7}\,$m, or alternatively $\lambda=10^{-6\pm2}$\,s$^{-1}$ at $\rC=10^{-6}\,$m, under the requirement of a collapse taking place in the mesoscopic regime during the process of latent image formation in photography}. 
On the other hand, there is a broad consensus in setting $\rC$ within the mesoscopic length scale of $\rC=10^{-7}$\,m.  This {choice} would guarantee microscopic superpositions to survive  and the suppression of {macroscopic ones, although only experiments can determine its value}. 

The DP model relates the collapse mechanism to gravity by choosing a correlator  proportional to the Newtonian potential $\mathcal{D}_\text{\tiny DP}({\bf x}-{\bf y})=\frac{G}{\hslash}\frac{1}{|{\bf x}-{\bf y}|}$, where $G$ is the gravitational constant. {When applying the model to a distribution of point-like particles the collapse rate  diverges, meaning that the collapse is instantaneous also for microscopic systems. This is clearly falsified by experimental evidence. For this reason, a}  regularization through the introduction of a spatial cutoff $R_0$ is needed, which gives a finite size to otherwise point particles. In the first formulation of the model, Di\'osi{~\cite{diosi1987universal}} suggested to set $R_0$ equal to the proton radius {of around} $10^{-15}$\,m, giving the model the particular appeal of being free from fitting parameters {and leading to a collapse time for a proton in a spatial superposition of around $10^6$ years, which is fully compatible with observations}. 
{However, as we will discuss in detail below, the spontaneous collapses induce an  increase of the mean energy of the system, resulting in a heating. Comparison with experimental data shows that $R_0=10^{-15}$\,m and smaller} values must be excluded as they would lead to an unphysical heating~\cite{ghirardi1990continuous}: the energy increase for a free nucleon would be of the order of $10^{-20}$\,erg/s for $R_0=10^{-15}$\,m, corresponding to a temperature increase of $7\times 10^{-5}$\,K/s{.} Over the life of the Universe{,}  a free nucleon would have developed a temperature of about $3\times10^{13}\,$K due to the DP noise, which is  not compatible with observations{~\cite{ghirardi1990continuous}.}
A different estimate for $R_0$ was given by Penrose~\cite{penrose2000wavefunction, penrose2014gravitization}, effectively equating {$R_0$} to the width of the wavefunction of the system. {This keeps} the model free from any  fitting parameter   {endowing} it  {with} a cutoff that explicitly depends on the system under scrutiny. {Following the most recent literature, we consider $R_0$ as a free parameter, whose value is eventually constrained by experiments.}

The collapse parameters are ultimately bounded by experiments. In what follows, we will review a number of them.
For the CSL model, such bounds constrain the maximum value that can be taken by $\lambda$ at given values of $\rC$ as --- according to Eq.~\eqref{CSLme} --- the collapse effect grows with the value of $\lambda$. 
Conversely, for the DP model, lower bounds on $R_0$ are sought, as the collapse strength depends inversely on {this} parameter.

Besides collapsing the system's wavefunction (or keeping it localized through time), the noise induced by  {the} collapse mechanism also results in  Brownian motion in addition to the system's dynamics. Detecting this motion is the goal of non-interferometric experiments.

\section{Interferometric and non-interferometric experiments}
\label{sec.interf}

The most direct approach for testing collapse models is to prepare a spatial quantum superposition, let the different components interfere --- ideally in a noise-free environment --- and then measure the corresponding interference pattern~\cite{arndt2014testing}.  If  interference fringes appear, the superposition principle holds for that type of systems within the measurement error, {otherwise it is violated. This can be due to different reasons, {such as} the localisation of the system's wavefunction predicted by a collapse model}.

{Interferometric} experiments face difficulties that limit their capability to place   experimental bounds on the collapse parameters. In particular, preparing and maintaining spatial superpositions of massive systems over time is challenging from a technical perspective as it requires   isolation from any external agent that might  spoil the superposition.  {Such an external action} would prevent the occurrence of possible collapse mechanisms or disguise them. Typically, this requires low temperature, high vacuum, low-vibration conditions.
Another major challenge is the experimental preparation of an initial coherent superposition state that is large enough to generate a visible interference pattern. The challenge of {the} preparation stage grows with the size and mass of the particles at hand. 
This {process} should be  robust {and} reproducible  as a large number of particles would need to be prepared in nearly identical initial states to allow for the acquisition of sufficient statistics to resolve an interference pattern.

State-of-art {interferometric} experiments  now employ particles of {around} $10^4$ atomic mass units (amu)  and have set an upper bound of $\lambda<10^{-7}$\,s$^{-1}$ at $\rC=10^{-7}$\,m for the CSL model \cite{fein2019quantum}. This is a few orders of magnitude away from testing Adler's value and is a billion times weaker than what is needed to probe the GRW value. Probing such a value would need masses of $10^7$\,amu and a size of the quantum superposition of {around} $180$\,nm, maintained for about $20${\,s}. 
The request on time is {too} demanding to make such {an} experiment practical. A potential way forward is to perform  {experiments} on-board of a dedicated satellite to exploit the advantages provided by {the} space environment~\cite{gasbarri2021testing,belenchia2021test}  {as its microgravity environment enables long free-fall times.} To date, interferometric experiments have not  set relevant bounds for the DP model. However, there are proposals for implementing experiments  that need  challenging technical developments{, mainly concerning how to generate spatial superpositions of massive systems}~\cite{marshall2003towards,machluf2013coherent,bateman2014near,howl2019exploring}.

{As} non-interferometric experiments do not rely upon the preparation of quantum superpositions,  {they provide}  an important advantage~\cite{collett2003wavefunction}. 
In fact, 
the collapse noise $W_t({\bf x})$ would act on the nucleons of a system regardless of the quantum or classical nature of the state it has been prepared into, making their dynamics stochastic. 
{The nucleons will randomly accelerate, {which} leads to a variety of effects  {that} will be discussed below. Among them,  a violation of the energy conservation principle {is predicted}. This should not be seen as disturbing in light of the phenomenological nature of the models being addressed.
}

As the typical strength of the collapse rate is very small, a successful experiment will still have to suppress other {noise} sources {from the environment}, as for the interferometric approach. The non-interferometric strategy is then to monitor the motion of a system in a controlled environment, looking for Brownian fluctuations, whose detection would be a first hint of a collapse effect. The lack of observation of such hints provides a bound on the collapse parameters, and allows {one} to draw 
so-called {exclusion plots} that identify the regions of parameters that need to be explored to rule out a given collapse model.  Figs.~\ref{fig:csl} and~\ref{fig:dp} report the exclusion plots for the CSL and the DP model, respectively. {Below we discuss individually the constraints obtained from the application of non-interferometric strategies.}

\subsection{Phonons in low temperature experiments}
\label{phonons}

The collapse noise affects  the collective dynamics of atoms, and modifies the phonon distribution in bulk materials,  leading to an increase of the internal energy of the system~\cite{bahrami2018testing,adler2018bulk}. The CSL model predicts a heating power  
given by
\bq \label{en_increase}
P_\text{\tiny CSL}=\frac{3}{4}\frac{\hslash^2 \lambda m}{m_0^2\rC^2},
\eq
where $m$ is the mass of the system. The system needs to be isolated in order to derive {significant} bounds.
{The main step in this direction is to perform the experiment at low temperatures, as in the case of the CUORE experiment~\cite{alduino2017projected}, where crystals of tellurium oxide weighting 340\,g are cooled to around 10\,mK. Also shielding the setup from other background noises --- such as $\gamma$ radiation or cosmic rays --- by resorting to underground facilities can improve the level of isolation~\cite{mishra2018testing}.}
Nevertheless, dissipative processes {due the interaction with the surrounding environment} 
will still take place. Therefore, materials with high density, which will enhance the collapse effect, and low thermal conductivity to reduce dissipation are the best candidates to test collapse-induced heating. 
Low-temperature experiments~\cite{pobell2007matter} can reach heating rates as low as  $P/m\sim 100$\,pW/kg. The most accurate modelling of energy deposition from radioactive decays and penetrating muons still leaves a residual heating of {around} $P/m\sim10$\,pW/kg 
unaccounted. This in turn sets the bound~\cite{adler2018bulk} $\lambda<3.3\times10^{-11}$\,s$^{-1}$ at $\rC=10^{-7}$\,m for the CSL model.

\subsection{Cold atoms}

State-of-the-art experiments in cold-atom technology allow {cooling} a cloud of atoms down to the pK scale, thus enabling a high degree of control of such systems.  {The} low operating temperature makes these systems  good candidates to test the effects of collapse models, although the amplification mechanism can not be exploited due to the negligible interaction among the atoms in the
cloud.

As for the system discussed in Sec.~\ref{phonons}, the collapse noise produces an increase of the energy (temperature) of the atoms in the cloud at a rate given in Eq.~\eqref{en_increase}, with $m$ now being the mass of an atom. 
Comparison with experimental data~\cite{laloe2014heating}, taking into account {several effects including} heating induced by three-body interactions or cooling resulting from evaporation, 
 leads to the upper bound   $\lambda<10^{-7\pm1}$\,s$^{-1}$ at $\rC=10^{-7}$\,m. A stronger bound can be obtained by considering diffusion in position~\cite{bilardello2016bounds}. CSL  predicts that the position variance grows as
\begin{equation}\label{x2t3}
    \braket{\hat {\bf x}^2}_t=\braket{\hat {\bf x}^2}^\text{\tiny QM}_t+\frac{\lambda\hslash^2}{2m_0^2\rC^2}t^3,
\end{equation}
where $\braket{\hat {\bf x}^2}^\text{\tiny QM}_t$ is the standard quantum mechanical spread, while the second term is the CSL-induced contribution. The latter grows as $t^3$, contrary to the linear increase of the CSL contribution to the energy. 
The ideal experiment to test this prediction  consists in cooling down an atomic cloud to very low temperatures, and then letting it  evolve {freely}. The CSL model predicts that the collapse noise will make the cloud expand faster than the predictions from quantum mechanics. If such {an} extra expansion is not observed, this can be used to set bounds on $\lambda$ and $\rC$.
Application of Eq.~\eqref{x2t3} to 
experimental data~\cite{kovachy2015matter}  {leads to} $\lambda<5.1 \times 10^{-8}$\,s$^{-1}$ {for the reference value of $\rC=10^{-7}$\,m}.   

\subsection{Optomechanical systems}

Optomechanical systems are based on the interaction between a mechanical oscillator and a radiation field shone on it~\cite{aspelmeyer2014cavity}. After the system has reached an equilibrium, one can infer the dynamical properties of the mechanical component, and consequently their modification due to external influences such as  those {caused by} collapse, by analyzing the radiation field~\cite{bahrami2014proposal,nimmrichter2014optomechanical,diosi2015testing}. The mechanical oscillator, which is driven by the radiation-pressure coupling with the radiation field, is assumed to be immersed in a thermal bath at temperature $T$, whose action is quantified by a temperature-dependent noise and a dissipation. The overall noisy action on the mechanical system is characterized in terms of the {density noise spectrum} of the oscillator’s position, which reads~\cite{bahrami2014proposal,nimmrichter2014optomechanical,diosi2015testing}:
\bq
S_\text{\tiny DNS}(\omega)=S_\text{opto}(\omega)+\frac{\hslash \omega m \gamma_\text{m} \coth(\hslash \omega/2\kB T)+S_\text{\tiny CM}}{m^2[(\omega_\text{eff}^2-\omega^2)^2+\gamma_\text{eff}^2\omega^2]},
\eq
where $S_\text{opto}(\omega)$ is the standard optomechanical contribution from the radiation field on the mechanical resonator {at frequency $\omega$}. The second term --- the contribution from the environment --- is characterized by the mass $m$ of the mechanical part, the mechanical damping $\gamma_\text{m}$, the effective frequency and damping $\omega_\text{eff}$ and $\gamma_\text{eff}$ respectively{, where $\kB$ is the Boltzmann constant}. 
Collapse models contribute to the expression of the  {density noise spectrum} 
with the addition of $S_\text{CM}$, which depends on the mass density of the system, is proportional to $\lambda$, and can be interpreted as a variation of the {equilibrium} temperature (or energy $E$) of the system. 
Equivalently, due to the equipartition theorem, an increase of the effective energy of the system is translated into an increase  of the spread in position~\cite{bahrami2014proposal,nimmrichter2014optomechanical,diosi2015testing} given by  $\braket{\hat x^2}\sim \int\D\omega S_\text{\tiny DNS}(\omega)\propto E+\Delta E_\text{\tiny CM}${, where $\Delta E_\text{\tiny CM}$ is the collapse models' contribution to energy}. 

Several experiments with optomechanical systems imposed significant  bounds on the collapse parameters; we can  {separate these experiment into} three main classes. The first class is that of clamped systems, {as cantilevers}, where the motion of a ferromagnetic sphere {that is} attached at the end of a silicon cantilever is {examined}  using a superconducting detector. The system {with masses}   from tens \cite{vinante2016upper} to hundreds \cite{vinante2017improved} ng is monitored at different temperatures from 10\,mK to 1\,K to characterize the collapse induced increase of the effective temperature.  {These} tests constrained {the} CSL {model} to {around $\lambda<1.9\times 10^{-8}$\,s$^{-1}$} at $\rC=10^{-7}$\,m. Recently,  the setup \cite{vinante2020narrowing} was specifically tailored for testing the CSL model at $\rC=10^{-7}$\,m~\cite{ferialdi2020continuous}{, yielding an} upper bound $\lambda<2.0\times 10^{-10}$\,s$^{-1}${, which completely} covers the values suggested by Adler. The second class of experiments includes the gravitational wave detectors LIGO~\cite{PhysRevLett.116.061102}, AURIGA~\cite{vinante2006present}, and the space-based prototype LISA Pathfinder~\cite{armano2016sub,armano2018beyond}. These employ macroscopic masses from the kg to the ton scale{,}  whose motion is monitored with optical techniques, making them effectively optomechanical systems. Although being fully in the classical, macroscopic regime, such experiments pose the strongest experimental bounds on the collapse parameters~\cite{carlesso2016experimental,helou2017lisa,carlesso2018non}  for $\rC>10^{-5}\,$m.  {This is due to the} fact {that} for such large masses, the collapse  is magnified due to the amplification mechanism{. Although } the corresponding bounds are the strongest for large values of correlation length $\rC$, they {are} softer at $\rC=10^{-7}$\,m{: LIGO~\cite{PhysRevLett.116.061102} sets} $\lambda<1.0\times 10^{-5}$\,s$^{-1}$, {AURIGA~\cite{vinante2006present} gives} $\lambda<4.6\times 10^{-2}$\,s$^{-1}${, while LISA Pathfinder~\cite{armano2016sub,armano2018beyond} provides}  $\lambda<3.8\times 10^{-9}$\,s$^{-1}$. The third class of experiments is that of levitated systems. The levitation of spheres of around 0.1 to 5\,pg was {made} possible  through the use of a linear Paul trap \cite{pontin2020ultranarrow} and a magneto-levitational trap \cite{zheng2020room} at room temperature{, respectively}.
{The current bounds obtained from such} experiments are comparable to those from interferometric experiments,  {yielding} $\lambda<4.1\times 10^{-5}$\,s$^{-1}$ and $\lambda<6.7\times 10^{-7}$\,s$^{-1}$ for {the} CSL {model} at $\rC=10^{-7}$\,m{. Although these bounds are not yet competitive compared to other non-interferometric methods,  they} hold the promise to  {provide stricter bounds}. One can expect a major improvement when working  {in cryogenic conditions}. 

\subsection{Gamma and X-ray emission} \label{sec.radiation}

Brownian motion, such as that induced by the collapse noise, imparts a (random) acceleration to particles, which makes them  radiate if charged{. As}  otherwise {this radiation} would not be there{, it} can be used 
to test collapse models. 

The most recent analysis applied to the CSL model~\cite{Donadi:2021tq} has shown that the radiation emission rate from a crystal is given by
\begin{equation}\label{eq:ratefinalevero}
\frac{\D\Gamma_\text{\tiny CSL}}{\D E}=N_\text{atoms}\ \frac{\left(N_{A}^{2}+N_{A}\right)\lambda \hslash e^{2}}{4\pi^{2}\varepsilon_{0}m_{0}^{2}\rC^{2}c^{3}E},
\end{equation}
where $N_\text{atoms}$ is the total number of atoms, $N_{A}$ the atomic number, {$e$ the elementary charge, $\epsilon_0$ the vacuum's dielectric constant, $c$ the speed of light} and $E$ the energy of the emitted photons. 
Eq.~\eqref{eq:ratefinalevero} is valid for $E\in[10, 10^5]$\,keV, which  corresponds to photon wavelengths larger than {the} size {of a nucleus} but smaller than {that of an}  {atom}.  {In this regime, the} protons in the same nucleus emit coherently{,}  giving {rise} the quadratic contribution $ N_{A}^2${. Because}  the electrons emit incoherently from the atomic nuclei,  their contribution {does not} cancel that of the protons {and} the electrons contribute linearly with $ N_{A}$. A similar expression is derived for the DP model~\cite{donadi2020underground}. 

The first application~\cite{diosi1993calculation} of  {the induced radiation emission rate} ruled out the Karolyazy model~\cite{karolyhazy1966gravitation}, which proposes a connection between wavefunction collapse and gravity. It was later applied to the mass-independent version of the CSL model{,} where the mass density $\hat M({\bf x})$ in Eq.~\eqref{csl} is replaced by the particle number density times $m_0$, effectively ruling it out~\cite{fu1997spontaneous}. 

A recent comparison with data from a dedicated experiment -- performed in the underground Gran Sasso laboratories in Italy  lead to the strongest bounds on {the} CSL~\cite{Donadi:2021tq} and on the DP~\cite{donadi2020underground} model {of} $\lambda<5.2 \times 10^{-13}$\,s$^{-1}$ at $\rC=10^{-7}$\,m  and {of} $R_0\ge0.54 \times 10^{-10}$\,m{, respectively}.  {It also} ruled out the parameter-free version of the DP model relating $R_0$ to the width of the wave function, as suggested by Penrose. According to this prescription, one would expect  
$R_0\sim 5 \times 10^{-12}$\,m{ for Germanium crystal cooled down to 77\,K}, which is about 10 times smaller than the lower bound set by the experiment.

\subsection{Decay of superconducting currents in SQUIDs}
Below a critical temperature, metals become superconductors: electrons bind in pairs, {so-}called Cooper pairs{,} and flow without resistance on the metal surface~\cite{Tinkham1996}.
A particularly interesting instance of  {such} devices is given by superconducting quantum interference devices ({known as} SQUIDs), which are characterized by a superconducting loop interrupted by two Josephson junctions. 
{It was suggested}~\cite{leggett1980macroscopic} {--- and later achieved\cite{friedman2000quantum} ---} that SQUIDs can be put in the superposition of two macroscopically dinstinct current states, and that these could be exploited to test the validity of the superposition principle. 

Collapse models predict that  {superconducting} currents are unstable, because the collapses tend to localize 
 single electrons, thus breaking Cooper pairs, leading to the decay of the current~\cite{rae1989can,buffa1995dissipation}. Such an effect is suppressed by the small value of the electron mass with respect to the nucleon reference mass, but is enhanced by the large number of electrons taking part  {in} the process. For the CSL {model}, the decay rate can be approximated  {as}~\cite{buffa1995dissipation}
 \bq
 \gamma_\text{\tiny CSL}=\frac{3}{2\sqrt{\pi}}\frac{N}{k_\text{F}}\frac{\lambda}{\rC},
 \eq
where $N$ is the number of Cooper pairs, and $k_\text{F}$ is the Fermi momentum.  
 This  {is} compared with the experimental rate{\cite{crowe1957trapped,Tinkham1996} $\gamma\sim 3\times10^{-13}$\,s$^{-1}$}, which is obtained by measuring  the  decay of the field produced by the  {superconducting currents~\cite{crowe1957trapped}}, which allows to set  {an} upper bound on the CSL rate {of}~\cite{adler2007lower} $\lambda<10^{-3}$\,s$^{-1}$ at $\rC=10^{-7}$\,m. The theoretical estimate of the supercurrent decay, however, neglects the recombination of electrons into Cooper pairs, thus the bound could be  weaker. However,  {because} the experimental data on  {superconducting currents} decay are  dated~\cite{crowe1957trapped}, more recent measurements could possibly allow to set stronger bounds. 

\subsection{Astronomical and cosmological observations}
\label{cosmology}

Astronomy and cosmology are becoming more and more important for testing collapse models, because they provide an arena{,} where the collapse effects can build up over very long times and for very large systems~\cite{adler2007lower}. 
In the non-relativistic regime, one can exploit the collapse-induced Brownian motion to set  bounds on the collapse parameters, which are reported in Table \ref{tab:cosmology}. 

The collapse noise reduces the stability of bound systems, and this can be applied to a variety of situations. {The} dissociation of cosmic hydrogen during the evolution of the Universe~\cite{PhysRevLett.73.1},  {results in} the bound $\lambda<1$\,s$^{-1}$ for $\rC=10^{-7}$\,m. The same noise, by accelerating protons, perturbs the thermal history of the Universe. {Besides the high energy photons considered in Section \ref{sec.radiation}, protons will also emit low energy photons,}  which  contribute to the Cosmic Microwave Background (CMB) radiation; precision measurements of the latter give~\cite{adler2007lower} $\lambda<10^{-5}$\,s$^{-1}$ for $\rC=10^{-7}$\,m.  { {Because} the emission {is} not thermal, these photons will} distort the spectrum of the CMB{. Data}  from the COBE/FIRAS (Cosmic Background Explorer/Far Infrared Absolute Spectrophotometer) observation{s} 
bounds the CSL parameters to~\cite{PhysRevD.86.065016} $\lambda<10^{-1}$\,s$^{-1}$ for $\rC=10^{-7}$\,m. 

The intergalactic medium,  {consisting} of highly ionized hydrogen,  is heated by various astrophysical sources and is cooled by adiabatic expansion of the Universe and by recombination cooling of the plasma. {As} the collapse noise will add to the heating mechanism,  {it will increase}  the equilibrium temperature. 
Observations set the bound~\cite{adler2007lower} $\lambda<10^{-8}$\,s$^{-1}$. 

Another equilibrium argument can be applied to astronomical and astrophysical bodies, such as {Neptune~\cite{adler2019testing} and} the neutron star~\cite{tilloy2019neutron} PSR J 1840-1419{, which is} the coldest neutron star found so far{. Under the assumption}  that the collapse-induced heating is equilibrated by the energy loss due to  the radiation emission, as described by the Stefan-Boltzmann law,  {one obtains} $\lambda<9.4\times10^{-7}$\,s$^{-1}$ for PSR J 1840-1419 and $\lambda<6.6\times10^{-11}$\,s$^{-1}$ for Neptune. 

\begin{table*}[t]
    \centering
    \begin{tabular}{c|c}
    \hline
        \textbf{Effect} & \textbf{Bound on $\lambda$} [s$^{-1}$]\\
        \hline
        \hline
         Non-dissociation of  hydrogen~\cite{PhysRevLett.73.1}& $<1$  \\
         CMB distorsion (COBE/FIRAS)~\cite{PhysRevD.86.065016}&$<10^{-1}$\\
        {Protons heating's contribution to the CMB}~\cite{adler2007lower}&$<10^{-5}$\\
         Heating in neutron stars~\cite{tilloy2019neutron} & $<9.4\times 10^{-7}$\\
         Heating of the Intergalactic Medium~\cite{adler2007lower}& $<10^{-8}$\\ 
         Heating of  Neptune~\cite{adler2019testing} & $<6.6\times 10^{-11}$\\
    \hline
    \end{tabular}
    \caption{{Astronomical and cosmological bounds on the CSL model. The listed bounds, which are discussed in Section~\ref{cosmology}, are computed for the reference value of the characteristic length $\rC=10^{-7}\,$m}. The strongest bound is reported also in Fig.~\ref{fig:csl}. }
    \label{tab:cosmology}
\end{table*}

Collapse models have {also} been applied  to cosmology
. They were proposed as candidates to implement an effective cosmological constant~\cite{PhysRevLett.118.021102}, or to justify the emergence of the cosmic structures in the Universe~\cite{perez2006quantum,PhysRevD.85.123001,PhysRevD.88.085020}, whose imprint can be found in the observed temperature anisotropies of the CMB. The latter {is} a remarkable prediction of inflationary cosmology, where theory and observations match very well.  {Collapse} dynamics {having acted}  since  {shortly} after the Big Bang will impact the spectrum of primordial perturbations, both at the scalar and {at the} tensorial level~\cite{PhysRevD.87.104024,PhysRevD.90.043503,leon2015inflation,PhysRevD.95.103518,PhysRevD.102.043515}.

{Under this perspective, o}bservational data {applied to cosmic inflation}  {were} used to rule out the CSL model for a specific choice of the relativistic collapse operator~\cite{martin2020cosmic}, but soon after it was shown that a different choice~\cite{gundhi2021impact} restores compatibility of CSL with cosmological observations. The problem  is that it is not clear  which form collapse models should take in relativistic situations{~\cite{Bengochea:2020vr}} {---} and even less in situations where gravitational effects are strong. 

\section{Perspectives}
To further progress in testing collapse models, new dedicated experiments will have to be designed and performed to achieve unprecedented levels of control over the relevant degrees of freedoms of the probe mass. They will push for  technological developments, which in turn will open  the possibility of discovering new physical properties. Here, we will review some promising avenues that are currently {being} explored.

The first possibility is to test collapse models using parametric heating of a trapped nanosphere. Specifically, a Paul trap is proposed~\cite{goldwater2016testing} to measure the heating rate of a single-charged levitated nanosphere.  {The} hybrid trap cools the mechanical motion to a low temperature{, and afterwards} the optical field of the cavity {is turned off} to let the nanosphere evolve freely before measuring the particle's energy. {By comparing}  the predictions with a model including the  heating induced by the collapse mechanism {one can test the} parameter range  to $\lambda=10^{-12}$\,s$^{-1}$ for a background pressure of $10^{-13}$\,mbar and a temperature of the mechanical system of $20$\,K.

Although being commonly the first candidate in many experiments, translational degrees of freedom are not the only available {option}. Indeed, it is possible to provide very stringent constraints {on} the collapse parameters by using roto-vibrational degrees of freedom. 
A master equation describing the roto-vibrational diffusion due to collapse effects has been derived~\cite{Schrinski2017}, which is used
in a non-interferometric proposal~\cite{carlesso2018non} applied to an optomechanical system. {Such proposal demonstrated that roto-vibrational diffusion can be employed}  to restrict the uncharted values of the collapse parameters using both lab-based and space-bound configurations{, potentially down to the GRW parameters}. 

Performing non-interferometric experiments in free-fall
is another possible way to enhance  the constraints on the collapse parameters~\cite{gasbarri2021testing}. Indeed, in free fall, the system does not require external potentials that would inevitably introduce extra noises in the system's dynamics, hindering those due to the collapse mechanism.  Concrete possibilities on ground are provided by the Bremen drop-tower~\cite{gierse2017fast} or the Hannover Einstein Elevator platform~\cite{lotz2020tests}.  {Such experiments could also be performed}  in dedicated space  missions~\cite{kaltenbaek2016macroscopic,gasbarri2021testing} or on board the International Space Station{,} where other quantum experiments {were} already {conducted}~\cite{elliott2018nasa}.

The {performance of} collapse models {test} can also be enhanced significantly by  {incorporating} information-theoretic techniques of sensing and metrology~\cite{parisIJQI}. In particular, building on the success in estimating the temperature of open quantum systems~\cite{Brunelli2011,Brunelli2012},  quantum parameter estimation techniques can be employed as a way to infer the equilibrium temperature of a mechanical oscillator potentially subjected to the effects of {the CSL model}.

One can complement the latter schemes with the use of hypothesis testing methods. By making use of Bayesian test protocols applied to both matter-wave interferometry~\cite{schrinski2020,schrinski2020rule} and non-interferometric settings~\cite{marchese2021}, one {can address} the hypothetical modifications of quantum theory induced by the occurrence of collapse mechanisms. 

{Current} state-of-art non-interferometric investigations {can be extended} to the possible {generalizations of collapse models. The CSL and DP models resort to a white noise, which is not physical, which moreover breaks the energy conservation of the system. The full resolution of both limitations requires the development of an underlying theory, which is not yet available, although some work in this direction has been made~\cite{Adler:2004wc}. Meanwhile,} non-white and dissipative generalizations of the CSL~\cite{adler2007collapse,smirne2015dissipative} and DP~\cite{PhysRevA.90.062105} models {have been formulated}. In the former extension, a cut-off frequency $\Omega_{0}$, a new collapse parameter, characterizes the noise spectrum, making it more similar to other physical noises. On the other hand, the dissipative extension avoids the energy  of an otherwise isolated system to diverge. In such a model, the system eventually thermalizes to a temperature $T_{0}$, which is a further collapse parameter. There are currently several experiments providing bounds on the collapse parameters of these extensions~\cite{nobakht2018unitary,carlesso2018colored,pontin2020ultranarrow,vinante2020testing}. However, with the additional parameters $\Omega_{0}$ and $T_{0}$, the parameter space widens, and thus  {it becomes more challenging to}   fully cover its unexplored regions.

{More ambitiously, collapse models call for an underlying deeper-level theory where the unitary dynamics, as well as the collapse, emerge naturally. This would explain the physical origin of the collapse of the wave function, be it related to gravity as suggested by Penrose~\cite{penrose1996gravity} and others --- or to  yet unidentified degrees of freedom~\cite{Adler:2004wc}.}

The interest in collapse models and their experimental testing has grown considerably in the last decade, {which is} also sustained by significant technological developments. The unprobed part of the parameter space  {has been greatly reduced}, pushing  the limits of quantum theory {further}. Nevertheless, the {question on whether}  quantum mechanics {is} valid universally up to the macroscopic scale {remains open.}  {And only}  experiments can tell.

\section*{Acknowledgments}
The authors acknowledge fruitful discussions with Roger Penrose and Andrea Vinante on various aspects of the models and the related experiments.
MC and MP are supported by UK EPSRC (Grant No.~EP/T028106/1).
SD and AB acknowledge financial support from INFN.
LF, MP, HU and AB acknowledge financial support from the H2020 FET Project TEQ (Grant No.~766900).
MP acknowledges the DfE-SFI Investigator Programme (Grant No.~15/IA/2864), the Leverhulme Trust Research Project Grant UltraQute (Grant No.~RGP-2018-266), the Royal Society Wolfson Research Fellowship scheme (Grant No.~RSWF\textbackslash R3\textbackslash183013) and International Mobility Programme.
HU acknowledges financial support from the Leverhulme Trust (Grant No.~RPG-2016-04), and EPSRC (Grant No.~EP/V000624/1).
AB acknowledges  the Foundational Questions Institute and Fetzer Franklin Fund, a donor advised fund of Silicon Valley Community Foundation (Grant No.~FQXi-RFP-CPW-2002) and the University of Trieste.

\section*{Competing  interests}
The authors declare no competing financial{/non-financial} interests.

\end{document}